\documentclass[letter,twocolumn]{jpsj3}
\usepackage{txfonts}
\usepackage{bm}

\title{Theory of the Strain Engineering of Graphene Nanoconstrictions}

\author{Masahiko Hayashi$^1$\thanks{m-hayashi@ed.akita-u.ac.jp}, Hideo Yoshioka$^2$, Hikari Tomori$^3$, Akinobu Kanda$^3$}
\inst{$^1$Faculty of Education and Human Studies, Akita University, 010-8502, Akita, Japan\\
$^2$Department of Physics, Nara Women's University, Nara 630-8506, Japan\\
$^3$Division of Physics, Faculty of Pure and Applied Sciences, University of Tsukuba, Tsukuba, Ibaraki 305-8571, Japan} 

\abst{
Strain engineering is one of the key technologies for using graphene as an electronic device: 
the strain-induced pseudo-gauge field reflects Dirac electrons, 
thus opening the so-called conduction gap. 
Since strain accumulates in constrictions, graphene nanoconstrictions can be a good platform for this technology. 
On the other hand, in the graphene nanoconstrictions, Fabry-P\'erot type quantum interference 
dominates the electrical conduction at low bias voltages. 
We argue that these two effects have different strain dependence; 
the pseudo-gauge field contribution is symmetric 
with respect to positive (tensile) and negative (compressive) strain, 
whereas the quantum interference is antisymmetric. 
As a result, a peculiar strain dependence of the conductance appears even at room temperatures.}

\begin{document}
\maketitle

Graphene has been attracting attention as a base for future electronic devices, 
such as the field-effect transistors (FETs)\cite{Novoselov:2004ita,Novoselov:2005es}. 
In device application, controlling electrical conduction is a critical challenge, 
and strain engineering is one of the promising solutions\cite{Pereira:2009cha,Guinea:2009fla,Guinea:2010ffa,Guinea:2012gqa,Bahamon:2013baa}. 
This technique is based on the feature that strain acts as a 
pseudo-gauge field (magnetic field), reflecting Dirac electrons in graphene. 
Various attempts to introduce strain into graphene have been made so far
\cite{Lee:2008ew,Levy:2010hl,Tomori:2011wl,ReserbatPlantey:2014hw,Choi:2015ev,Jiang:2017kj,Pacakova:2017kv,Zhang:2018eb,Hsu:2020iga} and it is notable that graphene can tolerate strain up to 20\%\cite{Lee:2008ew}. 
This shows the great potential of strain engineering in graphene. 
In this Letter, we investigate strain engineering in graphene nanoconstrictions.  
Since strain is known to accumulate in the constrictions\cite{timoshenko1951theory,landau1986theory}, 
graphene nanoconstrictions can be a promising platform for strain engineering. 

The nanoconstrictions are also interesting 
because they clearly show the quantum interference of the Dirac electrons\cite{MunozRojas:2006kda}.
Especially, Fabry-P\'erot type interference is studied theoretically 
\cite{Shytov:2008ca,RamezaniMasir:2010hw, Darancet:2009dw,Yang:2014da} and 
experimentally \cite{Allen:2017hh,Ahmad:2019kw}. 
How quantum interference affects strain engineering is an interesting question. 

\begin{figure}[t]
\begin{centering}
\includegraphics[clip,width=5cm]{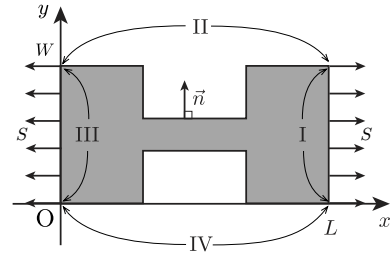} 
\par\end{centering}
\caption{Geometry of the sample. 
A strip is pulled in $x$-direction 
on two boundaries parallel to $y$-axis 
with a uniform stress $S$. 
Vector $\vec{n}$ is the normal vector to the boundary. 
The right, upper, left, and lower boundaries are labeled by I, II, III, and IV, respectively. }
\label{geometry} %
\vspace{-7mm}
\end{figure}

We study the effects of strain on 
the electrical conduction in graphene nanoconstrictions. 
The geometry is shown in Fig. \ref{geometry}. 
The graphene nanoconstriction 
is made of a single layer graphene and 
a uniform stress $S$ is applied parallel to the lengthwise direction of the wire. 
We consider both positive (tensile) and negative (compressive) stress 
for theoretical clarity, although the latter is difficult to realize experimentally in the present geometry. 
We start from the tight binding model of graphene, whose Hamiltonian is 
given by 
\begin{align}
H=- \sum_{\langle j,k\rangle}t_{jk}{c_{j\sigma}}^{\hspace{-1mm}\dagger} c_{k\sigma}+{\rm H.c}
\end{align}
where $c_{j\sigma}$ is the annihilation operator of an electron at the $j$-th site with 
spin $\sigma$. 
We limit the electron hopping to the nearest neighbor sites
$\langle j,k\rangle$. 
The hopping matrix element between the $j$-th and $k$-th site is denoted by $t_{jk}$, 
which is determined by taking into account the elastic deformation of the lattice due to stress. 

\begin{figure*}[h]
\begin{center}
\includegraphics[clip,width=17cm]{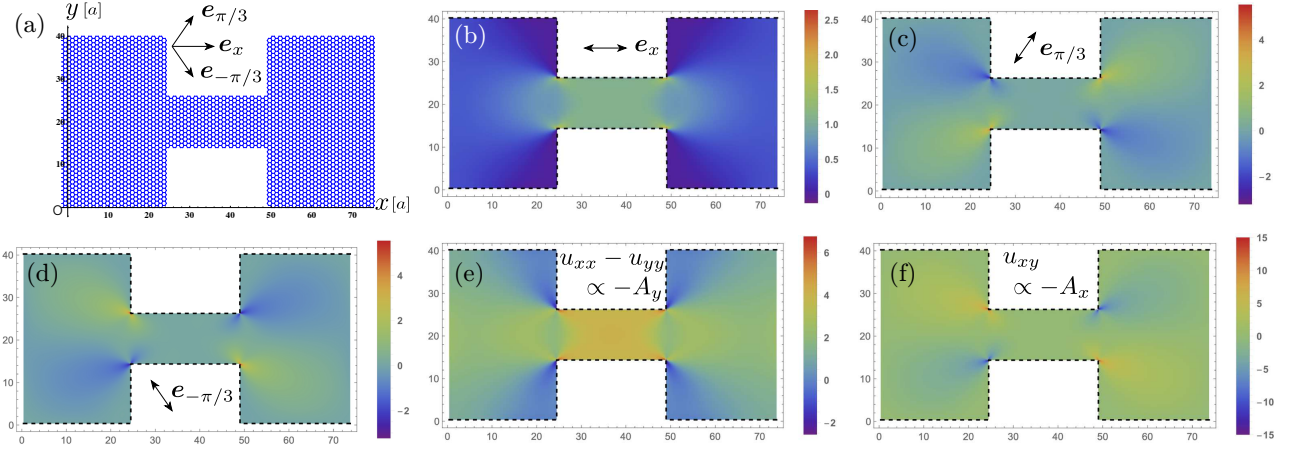} 
\end{center}
\caption{(color online) 
(a) Nanoconstriction considered in this Letter. Three direction of the bonds are shown. 
(b)$\sim$ (d) Color maps of the lattice distortion by a uniform stress $S$ (see Fig. \ref{geometry}), 
each corresponding to $\bm e_x$, $\bm e_{\pi/3}$, and $\bm e_{-\pi/3}$ direction, respectively. 
(e) and (f) are the plot of $u_{xx}-u_{yy}$ and $u_{xy}$ 
related to the pseudo-gauge field generated by strain. 
}
\label{strain} %
\vspace{-8mm}
\end{figure*}

First, we determine the strain distribution in the sample. 
We assume that graphene is an isotropic elastic media. 
Then, the stress tensor is expressed in terms of the stress function $\chi$ as 
\begin{align}
\sigma_{xx}&=\frac{\partial^2 \chi}{\partial y^2}, \,\,
\sigma_{yy}=\frac{\partial^2 \chi}{\partial x^2}, \,\,
\sigma_{xy}=-\frac{\partial^2\chi}{\partial x \partial y},  
\end{align}
and $\chi$ obeys the biharmonic equation $\Delta \Delta \chi =0$ with   
$\Delta$ being Laplacian\cite{timoshenko1951theory,landau1986theory}. 
The strain tensor $\hat{u}\equiv \{u_{\alpha\beta}\}$ ($\alpha, \beta =\{x, y\}$) is defined using 
displacement $\bm{u}(\bm{r})=(u_x,u_y)$ as 
\begin{align}
u_{\alpha\beta}=\frac{1}{2}
\left(\frac{\partial u_\alpha}{\partial \beta} +\frac{\partial u_\beta}{\partial \alpha}\right),
\end{align} 
and is related to the stress tensor by 
\begin{align}
u_{xx}&=\frac{\sigma_{xx}-\sigma \sigma_{yy}}{E}, \,
u_{yy}=\frac{\sigma_{yy}-\sigma \sigma_{xx}}{E}, \,
u_{xy}=\frac{1+\sigma}{E}\sigma_{xy},
\label{Hooke2}
\end{align}
where $E$ and $\sigma$ are Young modulus and Poisson ratio, respectively. 
The boundary conditions which $\chi$ satisfies are 
derived from the force balance equations at the boundaries. 
Let $\vec{n}$ be the unit vector normal to the boundary pointing outward 
as shown in Fig. \ref{geometry}, 
we set $\sum_\beta \sigma_{\alpha\beta}n_\beta=0$ 
on the boundaries II and IV, and $\sum_\beta \sigma_{x\beta}n_\beta= S$($-S$) on 
the boundary I (III). 
These are satisfied by setting as follows, 
\begin{align}
&\chi(0,y)=\chi(L,y)=\frac{S}{2}\left(y-\frac{W}{2}\right)^2,\,\,
\partial_x\chi(0,y)=\partial_x\chi(L,y)=0,\,\,\,
\nonumber\\
&\chi(x,y)|_{\rm II}=\frac{SW^2}{8}-\frac{SW}{2}y ,\,\,
\chi(x,y)|_{\rm IV}=\frac{SW}{2}y -\frac{3 SW^2}{8},
\nonumber\\
&\partial_x\chi(x,y)|_{\rm II}=\partial_x\chi(x,y)|_{\rm IV}=0,
\end{align}
where $\chi(x,y)|_{\rm II(IV)}$ is the value on the boundary II (IV). 
By discretizing the biharmonic equation, we solve $\chi$ numerically. 
Thereby we find the distribution of strain in the sample. 

The result corresponding to the stress $S/E=1$ is depicted in Fig. \ref{strain}. 
We can obtain the solution for $S/E\ne 1$ by multiplying a factor to $\chi$ 
because the equation that $\chi$ satisfies is linear. 
In Fig. \ref{strain}(a), we have shown the lattice and three unit vectors 
$\bm{e}_x$, $\bm{e}_{\pi/3}$ and $\bm{e}_{-\pi/3}$, 
corresponding to three directions of the atomic bonds. 
Here, the wire is aligned in the armchair direction 
so that the strain engineering works most efficiently\cite{Pereira:2009cha}. 
The change in the bond length between the $j$-th and $k$-th site is given, 
using strain tensor $\hat{u}$, 
as $|\bm{u}(\bm{R}_k)-\bm{u}(\bm{R}_k-\bm{\tau}_{jk})|\simeq
\kappa^{\,\,\,t}\bm{\tau}_{jk}\cdot \hat{u}(\bm{R}_k)\cdot\bm{\tau}_{jk}/b$, 
where $\bm{R}_k$ is the position of the $k$-th site, 
$\bm{\tau}_{jk}=\bm{R}_k-\bm{R}_j$, $b$ is the carbon-carbon distance (0.142 nm), and  
$\kappa$ is a numerical factor $\kappa\simeq 1/3$\cite{Suzuura:2002dz}. 
Figures \ref{strain}(b), (c), and (d) show the change rates of the bond length $\Delta b/b$ 
parallel to  $\bm{e}_x$, $\bm{e}_{\pi/3}$, and $\bm{e}_{-\pi/3}$, respectively. 
We see from Fig. \ref{strain}(b) that the bonds are significantly stretched 
in the constriction region as expected. 
From Figs. \ref{strain}(c) and (d), we see that the bonds at the corners 
are stretched (shrunk) depending on their direction. 
\begin{figure*}[h]
\begin{center}
\includegraphics[clip,width=17cm]{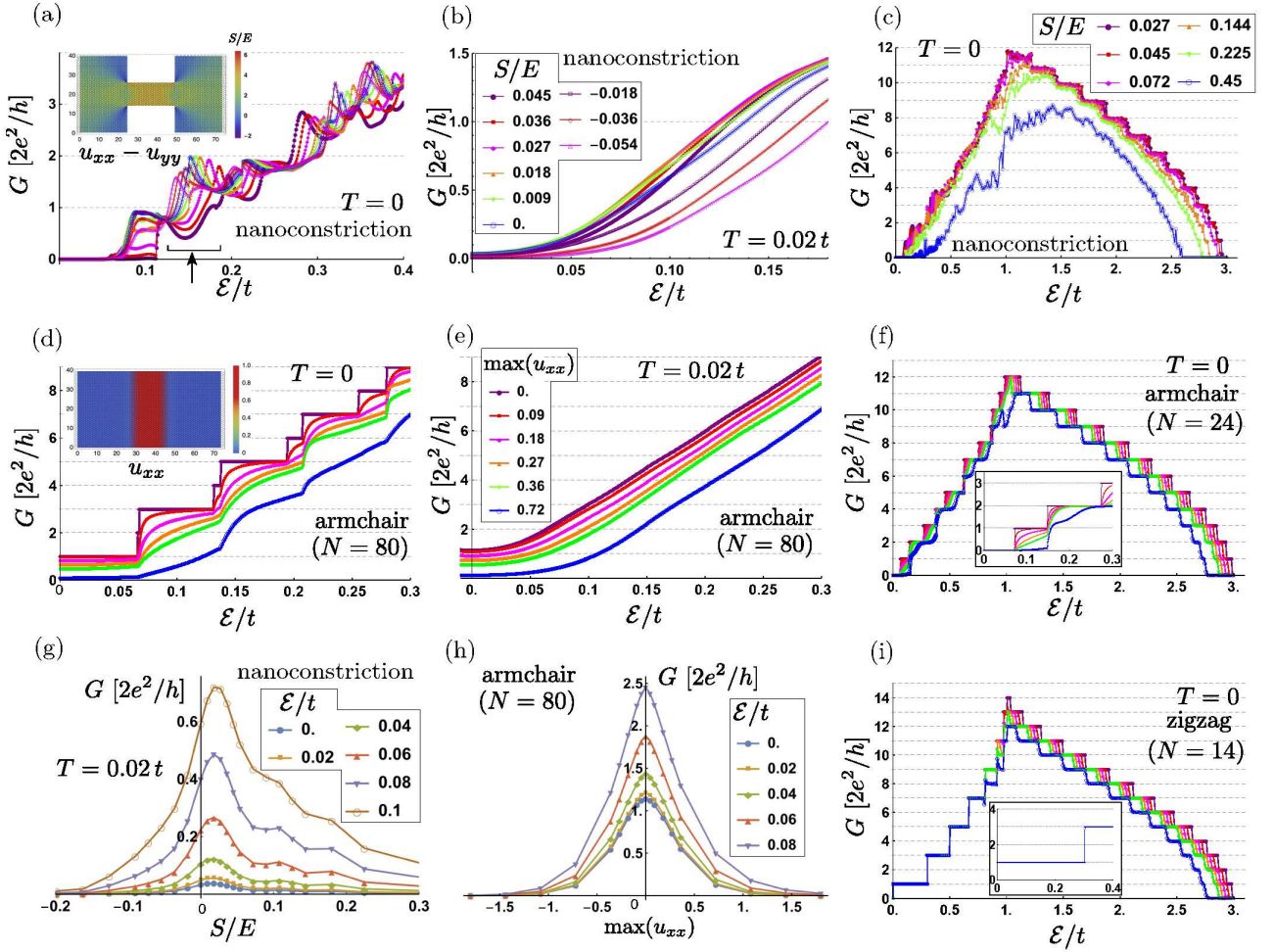} 
\end{center}
\caption{(color online) The conductance of the nanoconstrictions and nanoribbons: 
(a) $\sim$ (c) are the conductance of the nanoconstriction 
as a function of the energy $\cal{E}$ at different temperatures and energy ranges. 
The legend is common between (a) and (b). 
(d) and (e) are the conductance of the $N=80$ armchair nanoribbon. The legend is common. 
(g) and (h) are the stress (strain) dependence of the conductance for 
the nanoconstriction and $N=80$ armchair nanoribbon at different energies ${\cal E}$. 
(f) and (i) are the conductance of $N=24$ armchair nanoribbon and 
$N=14$ zigzag nanoribbon, respectively, whose legends are common with (e). 
The insets are the magnification of ${\cal E}\sim 0$ region. 
The conductance are symmetric with respect to ${\cal E}\rightarrow -{\cal E}$ in the present model and 
shown only ${\cal E}\ge 0$. As we see from Fig. \ref{strain}(e), $u_{xx}$ roughly corresponds to $4 \,S/E$ 
in the nanoconstriction. 
The sample geometries are shown in the insets of (a) and (d). }
\label{conductance} %
\vspace{-5mm}
\end{figure*}
In Figs. \ref{strain}(e) and (f) we have shown the profile of the pseudo-gauge field 
$\bm{A}=(A_x,A_y)$ 
\cite{Suzuura:2002dz,Manes:2007gca,Guinea:2009fla}. 
In our notation, $A_x=-2 \beta \kappa u_{xy}/a$ and 
$A_y=-\beta \kappa  (u_{xx}-u_{yy})/a$, 
where $\beta\simeq 2$ is a numerical factor, and $a=\sqrt{3}b$ is the lattice constant. 
We see that $A_y$, which is supposed to suppress the conduction along the wire, 
is largely generated at the constriction\cite{Pereira:2009cha}. 
We note that the pseudo-gauge field is not uniform in our system, and 
we also have $A_x$ as well as $A_y$. 
Their effects on the electrical conductance may be non-trivial. 

Base on the above results, we define the hopping matrix elements of the Hamiltonian $t_{jk}$ as 
\begin{align}
t_{jk}=t \exp\left(-\frac{\beta}{b^2}\,\kappa\,\,^{t}\bm{\tau}_{jk}\cdot \hat{u}(\bm{R}_k)\cdot\bm{\tau}_{jk}
\right),
\label{hopping-t}
\end{align}
where $t$ is the hopping integral without distortion 
\cite{Suzuura:2002dz,CastroNeto:2009cl}. 
Here we adopted exponential form so that it remains positive even for a large distortion. 

We have calculated the conductance of the wire $G({\cal E})$ at $T=0$ under strain, 
in terms of recursive Green function method\cite{Ando:1991wt}, 
where ${\cal E}$ is the energy of the incident electron or the bias voltage divided by 
the electron charge $e$. 
At a finite temperature $T$, the conductance $G({\cal E},T)$ is calculated from 
\begin{align}
G ({\cal E},T) &=\int_{-\infty}^\infty \frac{G({\cal E}')/(4 T)}{\cosh^2 (({\cal E}'-{\cal E})/(2T))} d{\cal E}'.
\end{align}

The wider and narrower part of the nanoconstriction 
are assumed to be $N=81$ and $N=25$ armchair nanoribbon, respectively\cite{Fujita:1996vs,N-def}. 
For comparison, we have also calculated the conductance of $N=80$ and $N=24$ armchair nanoribbons 
and $N=14$ zigzag nanoribbon, with an ideal uniaxial strain applied parallel to the lengthwise direction 
of the ribbon as depicted in the inset of Fig. \ref{conductance}(d). 

Figures \ref{conductance}(a) and (c) show the ${\cal E}$ dependence of the conductance 
of the nanoconstriction at $T=0$, and (b) is that at $T=0.02\, t$ 
($\sim$room temperature); (a)  and (b) are focused on 
the region near ${\cal E}=0$ whereas (c) shows the full band width. 
The legend is common between (a) and (b). 
We see from Figs. \ref{conductance}(a) and (c) 
that $G({\cal E})$ shows many peaks and jumps at low temperatures. 
We attribute this behavior to the Fabry-P\'erot type 
quantum interference of the Dirac electrons at the 
constriction\cite{Darancet:2009dw}, although 
the peaks in our results are much less periodic or sharp 
than those in a well-shaped Fabry-P\'erot interferometer\cite{width}. 
In a nanoconstriction made of a wire of length $L$, 
the Fabry-P\'erot resonance occurs 
when the characteristic wavelength $\lambda$ satisfies $2 L \sim n \lambda$, 
with $n$ being an integer. 
In the present case, $\lambda$ is the wavelength of the Bloch electron in the wire, 
and we may set $\lambda=h \,v_{\rm F}/{\cal E}=\sqrt{3}\pi\, a\, t/{\cal E}$  
with $v_{\rm F}$ being the Fermi velocity. 
If we put the effective wire length as $L/a\sim 25$, 
the resonant energy becomes ${\cal E}/t =\sqrt{3}\pi\, a\, n/(2 L) \sim 0.11 \,n$. 
This shows a rough match with the peak structure in Fig. \ref{conductance}(a). 

In Fig. \ref{conductance}(a), a peak located at ${\cal E}/t = 0.10 \sim 0.20$ 
(indicated by an arrow) shows a systematic shift to low bias side as the strain increases. 
The strain, which changes the hopping integral $t$, 
may shift the peak positions. 
Let us estimate its amount. 
From Fig. \ref{strain}(b), the change rate in the bond length at the constriction is 
$\Delta b/b=1.5 \sim 2.0$ for $S/E=1$ and, 
then, for $S/E=0.01$, $\Delta b/b$ is 1.5$\sim$2.0\%. 
According to Eq. (\ref{hopping-t}), the decrease in $t$ is estimated to be 3.0$\sim$4.0\%. 
Then, the peak position shift is expected to be the same amount, 
which roughly agrees with the peak shift observed in Fig. \ref{conductance}(a). 
For the understanding of the full properties, 
including the peak height and so on, we need more detailed analysis, however. 

The above features are in stark contrast to the case of armchair nanoribbons shown in 
Figs. \ref{conductance}(d) and (f), 
where the conductance steps are uniformly and monotonically smoothen by increasing strain. 

In Fig. \ref{conductance}(b), 
we see that, at a finite temperature, the conductance of the nanoconstriction 
shows no peaks or jumps, unlike $T=0$ case. 
However, it exhibits an irregular strain dependence compared to 
the armchair nanoribbon (Fig. \ref{conductance}(e)). 
In Fig. \ref{conductance}(g), we have plotted $G({\cal E},0.02\,t)$ for several 
values of $\cal{E}$ as a function of $S$. 
We see that the maximum of the conductance is located at a positive (tensile) strain side. 
Then, applying the stress at first increases conductance contrary 
to the strain engineering theory\cite{Pereira:2009cha}. 
In contrast, the conductance of $N=80$ armchair nanoribbon (Fig. \ref{conductance}(h))
has the maximum at the origin  
and is symmetric with respect to positive and negative $S$. 
The reason for this discrepancy could be attributed 
to the quantum interference effect which we have seen above for the $T=0$ case.
That is, the tensile stress pushes the peaks of quantum interference 
down to low energy side, and as a result, the low-bias conductance increases. 
This effect is antisymmetric with respect to $S$ in contrast to the 
pseudo-gauge field effect of strain engineering, which is symmetric\cite{symmetry}. 
In the present study, experimentally accessible region 
($u_{xx} \lesssim 20$\,\% or $S/E \lesssim 0.05$) 
covers only a narrow range of Figs. \ref{conductance}(g) and (h).  
We consider that our results may become easier to access by making the system size 
larger since the pseudo-magnetic field effect becomes more apparent. 

Figures \ref{conductance}(c), (f) and (i) show the conductance of 
the nanoconstriction, $N=24$ armchair nanoribbon and $N=14$ zigzag nanoribbon, respectively. 
In all the plots the strain effects are pronounced above ${\cal E}=1.0\, t$. 
However, we see that the behaviors near ${\cal E}=0$ are different; 
(c) shows resonance like peaks as we have seen in Fig. \ref{conductance}(a), 
(f) shows a monotonic decrease 
as strain is enhanced and (i) shows no dependence on the strain at all. 
The behavior of (f) and (i) are consistent with Ref. \citen{Pereira:2009cha}. 

From the above findings, we argue that the effects of strain on the conductance in 
nanoconstrictions are dominated by the two effects: 
1) {\it the pseudo-gauge field effect, which suppresses the conductance with stress-even dependence}, and 
2) {\it the Fabry-P\'erot type quantum interference effect, which enables resonant transport through the constriction, 
thus adding stress-odd contribution. }

Recently, several attempts to tune the stress on nanoconductors have been made, 
which enables highly controllable strain devices \cite{Androulidakis:2015iea,Caneva:2018kc}. 
We hope that our results may be useful for future experiments. 

In conclusion, we have studied the strain-dependent conductance of the 
graphene nanoconstrictions and found that interplay between 
pseudo-gauge field effect, which is expected from strain engineering theory, 
and Fabry-P\'erot type quantum interference effect, which is characteristic to 
nanoconstrictions, induces peculiar strain dependence of the electric 
conductance. 
Our results may be experimentally accessible. 

\vspace{1mm}
\begin{acknowledgment}
This work was supported by JSPS KAKENHI Grant Numbers 24540392 and 15K04619. 
\end{acknowledgment}


\begin{thebibliography}{10}
\bibitem{symmetry}
Here we note that the conductance depends on the constriction's detailed shape
  and the conductance maximum can be located in the negative $S$ region in
  other shapes. This is probably due to the complicated interference effects
  and will be discussed elsewhere.
\end{thebibliography}
\end{document}